# Confirmation of the monoclinic Cc space group for the ground state phase of Pb(Zr$_{0.525}$Ti$_{0.475}$)O$_3$ (PZT525): A Combined Synchrotron X-Ray and Neutron Powder Diffraction Study

Ravindra Singh Solanki[1], Sunil Kumar Mishra[2], Anatoliy Senyshyn[3], Songhak Yoon[4], Sunggi Baik[4], Namsoo Shin[5] and Dhananjai Pandey[1]

[1]*School of Materials Science and Technology, Indian Institute of Technology (Banaras Hindu University), Varanasi-221005, India*

[2]*Research and Technology Development Centre, Sharda University, Greater Noida-201306, India*

[3]*Forschungsneutronenquelle Heinz Maier-Leibnitz (FRM II), Technische Universität München, Lichtenbergstrasse 1, D-85747 Garching bei München, Germany*

[4]*Department of Materials Science and Engineering, Pohang University of Science and Technology, Pohang 790-784, Korea*

[5]*Pohang Accelerator Laboratory, Pohang University of Science and Technology, Pohang 790-784, Korea*

## Abstract

The low temperature antiferrodistortive phase transition in a pseudo-tetragonal composition of PZT with x=0.525 is investigated through a combined synchrotron x-ray and neutron powder diffraction study. It is shown that the superlattice peaks cannot be correctly accounted for in the Rietveld refinement using R3c or R3c+Cm structural models, whereas the Cc space group gives excellent fits to the superlattice peaks as well as to the perovskite peaks. This settles at rest the existing controversies about the structure of the ground state phase of PZT in the MPB region.

Lead Zirconate Titanate, $Pb(Zr_xTi_{1-x})O_3$ (PZT), is commercially used as a piezoelectric sensor and actuator material[1] due to its very high value of piezoelectric coefficients in the vicinity of a morphotropic phase boundary (MPB) at x≃0.520 in the phase diagram. The various stable phases of PZT at room temperature correspond to tetragonal[1] (space group P4mm), pseudotetragonal monoclinic[2, 3] (space group Cm), pseudorhombohedral monoclinic-I[2] (space group Cm), rhombohedral[1] (space group R3c) or pseudorhombohedral monoclinic-II[2] (space group Cc) and orthorhombic[1] (space group Pbam) structures stable for compositions with x≲0.515, 0.520≲x≲0.530, 0.530<x≲0.620, 0.620<x≲0.940 and x≳0.940, respectively. The MPB separates the stability fields of tetragonal[1] and pseudorhombohedral monoclinic phases through a thin region of a pseudotetragonal monoclinic phase[2]. The pseudotetragonal monoclinic phase of PZT was discovered by Noheda et al[3] for x=0.520 in a low temperature synchrotron X-ray diffraction (SXRD) study. They presented evidence for a tetragonal to monoclinic phase transition at T~250K. It is now known that this pseudotetragonal monoclinic phase gets stabilized at room temperature for x=0.525[4, 5]. Soon after the discovery by Noheda et al, Ragini et al[6] and Ranjan et al[7] presented evidence for an antiferrodistortive (AFD) phase transition at T≃210K using low temperature electron diffraction, neutron diffraction, dielectric and elastic constant measurements for the same PZT composition (x=0.520). This AFD transition is accompanied with unit cell doubling caused by antiphase rotation of oxygen octahedra in the neighbouring unit cells leading to the appearance of superlattice reflections which are visible in electron and neutron diffraction patterns only. Noheda et al[3] missed this superlattice phase in their SXRD studies as the superlattice reflections are not discernible in SXRD patterns due to weaker sensitivity of X-rays to small changes in oxygen positions caused by antiphase tilting of oxygen octahedra at the AFD transition. The space group of this superlattice phase was soon shown to be Cc by Hatch et al[8] and confirmed later on by Ranjan et al[9] and Noheda et al[10] who also showed that this phase coexists with the monoclinic Cm phase down to the lowest temperatures well below the AFD transition temperature for x=0.520. The existence of the low-temperature superlattice phase in the Cc space group was confirmed by other workers also in their transmission electron microscopy (TEM)[11,12] and neutron powder diffraction studies as a function of temperature[10,13] and pressure[13] and was nearly accepted as the true ground state of PZT. However some authors reported that the true ground state of PZT in and around the MPB is rhombohedral in the R3c space group and it coexists with the monoclinic phase in Cm space group over a wide composition range[14-20]. Thus the true symmetry of the low temperature phase of PZT has become controversial as evidenced by several recent publications on this system[17-23]. The main objection against the Cc space group is the non-observation of the $(1/2\ 1/2\ 1/2)_{pc}$ (pc stands for pseudocubic indices) superlattice peak which is permitted by the Cc space group but is extinguished in the R3c space group. However, the non-observation of the weak $(1/2\ 1/2\ 1/2)_{pc}$ superlattice peak in the neutron powder diffraction patterns is as enigmatic as the non-observation of the rather intense neutron $(3/2\ 1/2\ 1/2)_{pc}$, $(3/2\ 3/2\ 1/2)_{pc}$ and $(5/2\ 1/2\ 1/2)_{pc}$ superlattice peaks in the synchrotron X-ray powder diffraction patterns. Most likely, it is because the intensity of these reflections is too small[24] ($\lesssim 0.1\%$ of the most intense $(110)_{pc}$ peak) to be observable above the background counts for both the neutrons and the X-rays. In view of this, we recently[24,25] adopted an alternative strategy for resolving the Cc versus R3c controversy for the ground state of PZT. We analyzed the profile shape of the intense $(3/2\ 1/2\ 1/2)_{pc}$ peak which is a singlet for the R3c space group but a multiplet for the Cc space group. In order to be able to analyse the profile shape of this superlattice peak, we used higher wavelength neutrons ($\lambda$~2.5Å) to increase the angular separation of the $(3/2\ 1/2\ 1/2)_{pc}$ superlattice profile from the neighbouring intense $(111)_{pc}$ perovskite peak and we also replaced 6% $Pb^{2+}$ by a smaller cation $Sr^{2+}$ to increase the antiphase tilt angle and, hence, the intensity of the superlattice peaks. Using these two innovations, we could convincingly reject the R3c space group in a pseudotetragonal $(Pb_{0.06}Sr_{0.94})(Zr_{0.530}Ti_{0.470})O_3$ (PSZT530) composition and confirm the Cc space group[25] as the true ground state, as we could demonstrate that the $(3/2\ 1/2\ 1/2)_{pc}$ peak is not a singlet in PSZT530. However, since we substituted $Pb^{2+}$ with 6% $Sr^{2+}$ in these PZT samples, question arises about the validity of the conclusions arrived at for such modified PZT compositions to pure PZT compositions.

The motivation behind the present work is to settle the space group of the ground state of pure PZT (without $Sr^{2+}$ substitution at $Pb^{2+}$ site) by analyzing the profile shape of the $(3/2\ 1/2\ 1/2)_{pc}$ superlattice peak recorded using 2.536Å neutrons. For this, we have selected a PZT composition with x=0.525 i.e, PZT525, a composition that has been shown to possess pseudotetragonal monoclinic structure with Cm space group at room temperature[4, 5]. We present here results of Rietveld analysis of low temperature SXRD and neutron powder diffraction measurements on PZT525 that confirm the Cc space group and discard the R3c space group for the ground state phase of pure PZT in the MPB region.

Our samples were prepared by a semiwet route[26] which gives the narrowest composition width of the MPB region (0.515<x<0.530). Neutron powder diffraction (NPD) experiments (λ=2.536 Å) were carried out on high resolution SPODI powder diffractometer [27] at FRMII, Garching. High resolution synchrotron X-ray diffraction data were obtained from 8C2 HRPD beamline at Pohang Light Source (PLS), Pohang Accelerator Laboratory, Pohang at wavelength of 1.543 Å. Rietveld analysis of the neutron and synchrotron X-ray powder diffraction data was carried out using FULLPROF[28] software package.

Fig. 1 depicts the evolution of SXRD profiles of, $(200)_{pc}$ and $(220)_{pc}$ perovskite reflections of PZT525 with temperature in the range 10-700K. At 10K, the $(111)_{pc}$ perovskite peak is obviously not a singlet while the $(200)_{pc}$ peak profile shows large asymmetry indicating its non-singlet character as well. For a rhombohedral structure, $(111)_{pc}$ is a doublet but $(200)_{pc}$ is a singlet, whereas for tetragonal structure, the $(111)_{pc}$ is a singlet and $(200)_{pc}$ is a doublet. The observed features in Fig.1 at 10K rule out both these structures. As shown later on using Rietveld refinements, monoclinic phase in the Cm space group, discovered by Noheda et al[3] for x=0.520 just below the room temperature, can explain the entire diffraction pattern. As the temperature increases, there are distinct changes in the diffraction profiles and around 300K, new peaks marked with arrows appear in the $(200)_{pc}$ and $(220)_{pc}$ profiles. On increasing the temperature to 570K, the $(111)_{pc}$ peak becomes a singlet while $(200)_{pc}$ and $(220)_{pc}$ are doublet. This is the characteristics of a tetragonal phase. In the temperature range 300K≲T<570K, the tetragonal phase coexists with the monoclinic phase due to first order nature of the monoclinic to tetragonal phase transition[29]. At 700K, all the reflections are singlet as expected for the cubic paraelectric phase.

As is well known, XRD technique is not sensitive to small displacements of oxygen atoms due to the low atomic number. In order to capture the superlattice reflections resulting from the AFD transition in PZT involving small changes in oxygen positions, electron and neutron diffraction data are required, as was pointed out by Ragini et al[6] and Ranjan et al[7]. With a view to capture the low temperature AFD transition in PZT525, we depict in Fig. 2 the evolution of two such superlattice reflections with pseudocubic (pc) indices $(3/2\ 1/2\ 1/2)_{pc}$ and $(3/2\ 3/2\ 1/2)_{pc}$ with temperature. The presence of these superlattice reflections for T<225K confirms that the room temperature monoclinic phase in Cm space group of PZT525 undergoes an AFD transition into a superlattice phase, which is not revealed by the SXRD data at 10K shown in Fig. 1. As the temperature is increased form 4K, the intensity of the superlattice reflections decreases and becomes almost zero for T$\gtrsim$225K suggesting an AFD transition temperature of 200K<T<225K.

Since the superlattice reflections due to the AFD transition are not discernible in the SXRD patterns, the Cc and R3c space groups for the ground state phase of PZT525 become effectively equivalent to Cm and R3m space group. Accordingly, we carried out Rietveld analysis of the low temperature (10K) SXRD data using R3m, Cm and R3m+Cm space groups and the results of refinement are shown in Fig. 3. It is evident from this figure that the R3m model cannot explain the observed profiles as it leads to huge misfits and rather large $\chi^2$ values. On the other hand, refinement with single monoclinic phase in the Cm space group gives excellent fit between the observed and calculated profiles, as can be seen from Fig. 3(b). The value of $\chi^2$ dropped from 6.74 for R3m space group to 1.75 for the Cm space group. We also considered the possibility of coexisting R3c phase (which becomes R3m in the absence of superlattice peaks) along with Cm phase as proposed by several authors[14-20] for the ground state of PZT. Refinement using R3m+Cm coexisting model was carried out in three different ways. Firstly, we add a coexisting Cm phase to the refined structure obtained using single phase R3m space group. The value of $\chi^2$ reduced drastically (from 6.74 to 1.73) and the Rietveld fit improved. However, the phase fraction for R3m phase is reduced to ~10%. Secondly, we added the coexisting R3m phase to the refined structure obtained using single phase monoclinic Cm space group. In this case, phase fraction of R3m phase was less than 1% with little improvement in $\chi^2$. Thirdly, we refined the structure considering R3m and Cm phases simultaneously from the begining. In this case also phase fraction of R3m phase came out to be ~10%. The fits for the R3m+Cm model for the third strategy are shown in Fig. 3(c). The value of $\chi^2$ in all the three refinement is nearly same (~1.75) and also there is no significant difference in the quality of fits using the three two phase models. In view of this, the Cm space group model has to be preferred over the two phase R3m+Cm structural models since the number of refineable parameters is much less (24) than that for the two phase (37) models for a nearly comparable $\chi^2$ value. Thus, we conclude that our sample of PZT525 is nearly single phase monoclinic at 10K and at the SXRD level its space group is Cm. The results of Rietveld refinements above room temperature are not discussed here as these are already reported in a previous work[29]. In that work also, using Rietveld refinement technique, it was confirmed that the structure of PZT for x=0.525 is pseudotetragonal monoclinic in the Cm space group at room temperature. The present Rietveld refinement results for the same composition but at 10K neither support a Cm to R3c/R3m nor Cm to R3c+Cm/R3m+Cm AFD transition below room temperature (10K) proposed in references 14, 16 and 18.

To determine the symmetry of the low temperature phase resulting from the AFD transition, we carried out Rietveld analysis of the neutron powder diffraction data at 4K using the (i) R3c[1], (ii) R3c+Cm[14,16,18] and (iii) Cc[8-10] structural models. We have used anisotropic strain broadening parameters for all the three models. Due to the presence of superlattice reflections in the neutron powder diffraction pattern we first carried out the refinement using R3c space group but this single phase model leads to a large value of $\chi^2$ (~8.20) and huge misfit between observed and calculated patterns. We therefore added Cm phase in the refinement, as proposed by previous authors[14,16] and this reduced the phase fraction of the R3c phase to ~10% which is in agreement with the phase fraction of the R3m phase obtained from SXRD data. The overall fit corresponding to the R3c+Cm structural model is shown in Fig. 4(a). As is evident from the fits for the $(111)_{pc}$ and $(200)_{pc}$ perovskite peak shown in the inset of this figure, this model accounts fairly well for the perovskite reflections. However, the Cc space group model, which has only 30 refinable parameters as against 37 parameters for the R3c+Cm model, gives equally good, albeit marginally better, fit as can be seen from Fig. 4(b).

A correct structural model representing the true ground state of PZT should account not only for the main perovskite peaks but also the superlattice peaks. Surprisingly, in several recent studies[16, 18] carried out using R3c+Cm structural model, the fits corresponding to the superlattice reflections have not been shown for the R3c space group, the superlattice reflections are singlet but not for the Cc space group. Fig. 5 compares the Rietveld fits for the R3c+Cm and Cc models for the $(3/2\ 1/2\ 1/2)_{pc}$, $(3/2\ 3/2$ 1/2)$_{pc}$ and (5/2 1/2 1/2)$_{pc}$ superlattice peaks. It is evident from Fig. 5(a) that the R3c+Cm model gives huge misfit on the lower angle side of the (3/2 1/2 1/2)$_{pc}$ reflection and a misfit of 0.20$^0$ between the calculated and observed peak positions of the (3/2 3/2 1/2)$_{pc}$ reflection. Accordingly, the R3c+Cm model can be rejected out rightly. In literature[30, 31], the I4cm space group has also been proposed as the ground state phase but even this model fails to account for the superlattice peaks correctly as can be seen from Fig. 5(b). The Cc model on the other hand, gives an excellent fit to both the superlattice peaks as can be seen from Fig. 5(c). However, for the Cc space group, the (1/2 1/2 1/2)$_{pc}$ superlattice peak is not extinguished unlike that for the R3c space group. Fig. 5 gives fits near the (1/2 1/2 1/2)$_{pc}$ peak also. It is evident from the fitted curve shown in Fig. 5(c) that the intensity of the (1/2 1/2 1/2)$_{pc}$ peak is too small to be observable above the background counts/fluctuations. In fact, the calculated intensity of this peak is only ~0.12% of the strongest (111)$_{pc}$ perovskite reflection, which is of the order of the background counts in our neutron diffraction data[24]. It is important to emphasize that the non observation of (1/2 1/2 1/2)$_{pc}$ superlattice peak in the neutron diffraction pattern is as enigmatic as the non observation of (3/2 1/2 1/2)$_{pc}$, (3/2 3/2 1/2)$_{pc}$ and (5/2 1/2 1/2)$_{pc}$ superlattice reflections in the SXRD patterns, as the SXRD intensities of these peaks is comparable to the neutron intensity of the (1/2 1/2 1/2)$_{pc}$ peak. We simulated the SXRD pattern fixing the lattice parameters and coordinates of Table I and the intensity of the strongest superlattice peak (3/2 1/2 1/2)$_{pc}$ comes out to be ~0.08% of the intense (100)$_{pc}$ perovskite peak which is also within the background counts of the SXRD data. Further, if one considers only the value of $\chi^2$, one may arrive at an erroneous conclusion as the contribution of the weak superlattice peaks to the overall $\chi^2$ value is rather small. Our Rietveld fits for the other superlattice peaks clearly reject the R3c+Cm and I4cm+Cm models and favour the Cc space group. Table I lists the refined structural parameters for the Cc space group model at 4K. For the sake of completeness, we also list in Table I of the supplementary file[32] the refined coordinates for the R3c+Cm model at 4K.

To summarize, we have shown that the structure of the ground state phase of PZT525 does not correspond to the Cc space group in agreement with the results of the previous studies on pure as well as $Sr^{2+}$ substituted samples[24, 25], predictions of the first principles calculations[30] and high pressure studies[13]. In view of these findings, the recent claims of the R3c space group representing the structure of the ground state phase of PZT may be rejected, at least in the MPB region.

R. S. Solanki acknowledges financial support from Council of Scientific and Industrial Research (CSIR), India in the form of a Junior Research Fellowship. The work at Pohang Accelerator Laboratory has been supported financially by the Ministry of Science and Technology and Brain Korea 21 Program. The experiments at Pohang Light Source (PLS) were also supported by MOST and POSTECH, Pohang, Korea.

**Figure Captions:**

**Figure 1.** Temperature evolution of the synchrotron X-ray powder diffraction profiles of $(111)_{pc}$, $(200)_{pc}$ and $(220)_{pc}$ peaks of PZT525.

**Figure 2.** Temperature evolution of the superlattice $(3/2\ 1/2\ 1/2)_{pc}$ and $(3/2\ 3/2\ 1/2)_{pc}$ peaks in the neutron powder diffraction pattern showing an AFD transition at $200K<T_c<225K$.

**Figure 3.** The observed (dots), calculated (continuous line) and difference (bottom) SXRD profiles after Rietveld refinement of the structure at 10K using (a) R3m, (b) Cm and (c) R3m+Cm space groups. The insets show the enlarged views for the $(111)_{pc}$, $(200)_{pc}$ and $(220)_{pc}$ peaks.

**Figure 4.** The observed (dots), calculated (continuous line) and difference (bottom) neutron diffraction profiles after Rietveld refinement of the structure at 4K using (a) R3c+Cm and (b) Cc space groups. The insets show the fits for the $(111)_{pc}$ and $(200)_{pc}$ perovskite peaks.

**Figure 5.** The observed (dots), calculated (continuous line) and difference (bottom) neutron diffraction profiles of $(1/2\ 1/2\ 1/2)_{pc}$, $(3/2\ 1/2\ 1/2)_{pc}$, $(3/2\ 3/2\ 1/2)_{pc}$ and $(5/2\ 1/2\ 1/2)_{pc}$ superlattice peaks after Rietveld refinement of the structure at 4K using (a) R3c+Cm, (b) I4cm+Cm and (c) Cc space groups.

## Table Caption:

**Table I.** Refined structural parameters and agreement factors for PZT525 at 4K for the Cc space group

**Table I**

| Atoms | x | y | z | $B_{iso}(Å^2)$ |
|---|---|---|---|---|
| Pb | 0.00 | 0.75 | 0.00 | 1.13(4) |
| Ti/Zr | 0.219(5) | 0.253(6) | 0.247(3) | 0.06(3) |
| O1 | -0.042(3) | 0.253(2) | 0.011(2) | 1.05(3) |
| O2 | 0.207(5) | 0.513(7) | 0.017(2) | 0.03(1) |
| O3 | 0.186(3) | 0.004(1) | 0.479(1) | 0.46(2) |

a=10.0447(4)Å, b= 5.7335(2)Å, c= 5.76015(2)Å, β= 125.564°(2)

$\chi^2$= 1.85, $R_p$ =4.06, $R_{wp}$ = 5.15, $R_{exp}$=3.79

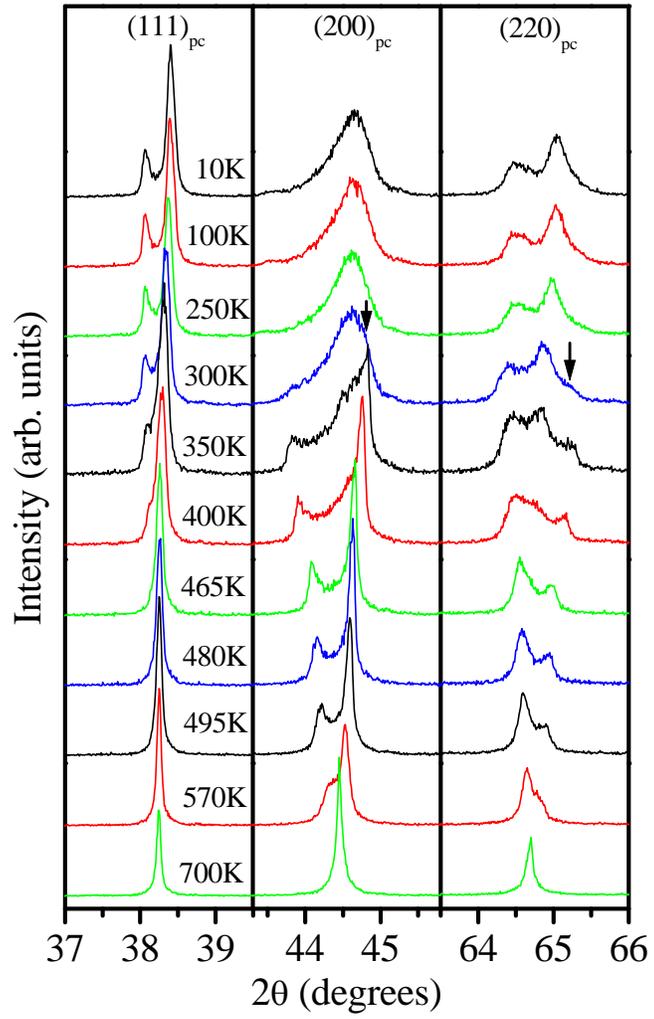

Fig. 1

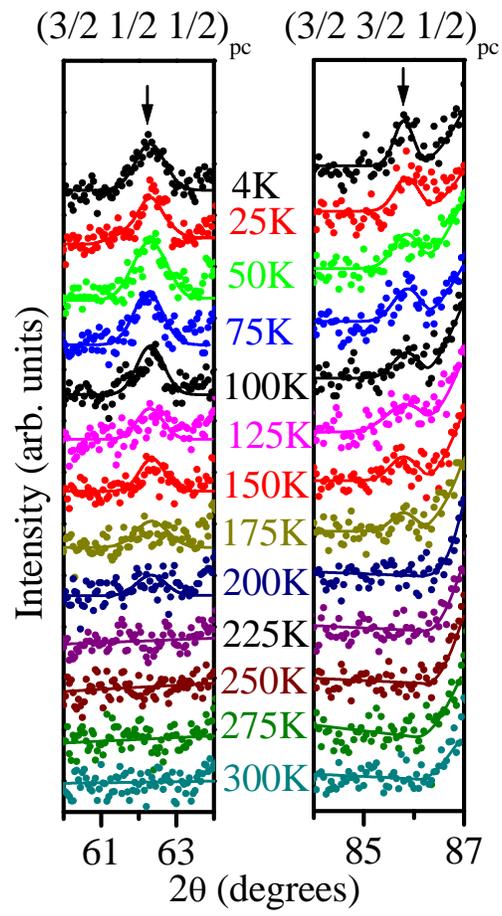

Fig. 2

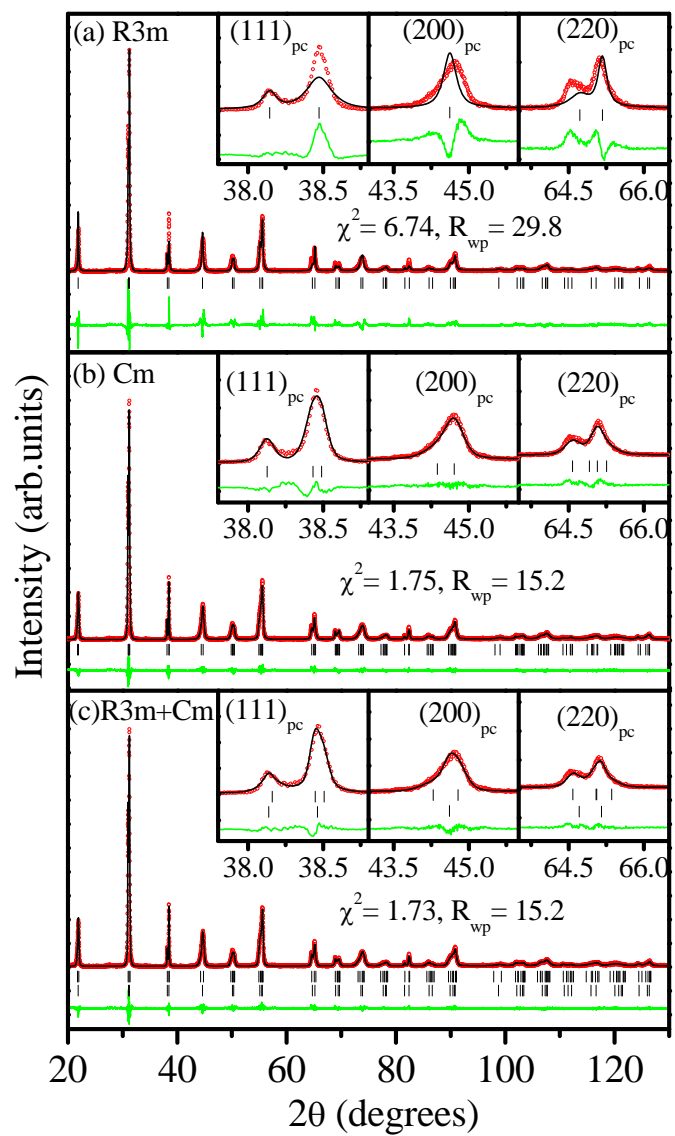

Fig. 3

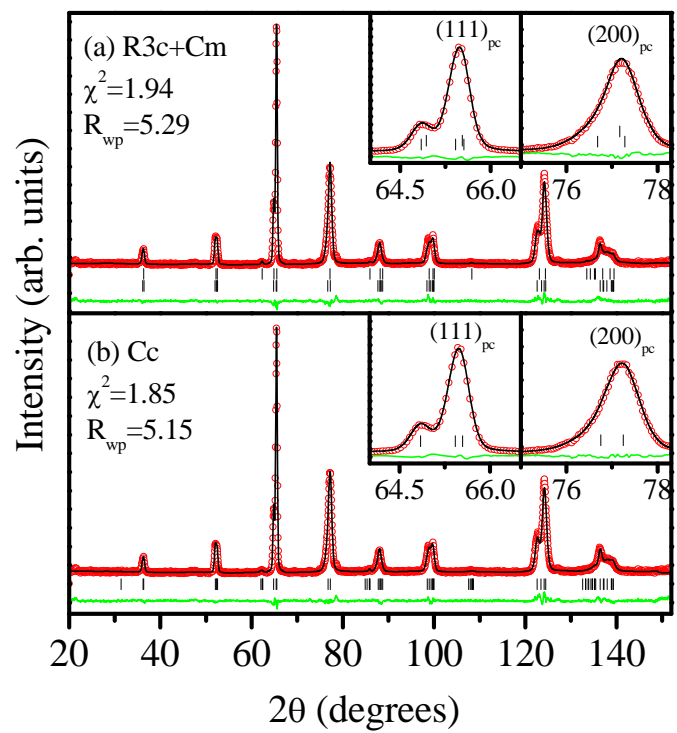

Fig. 4

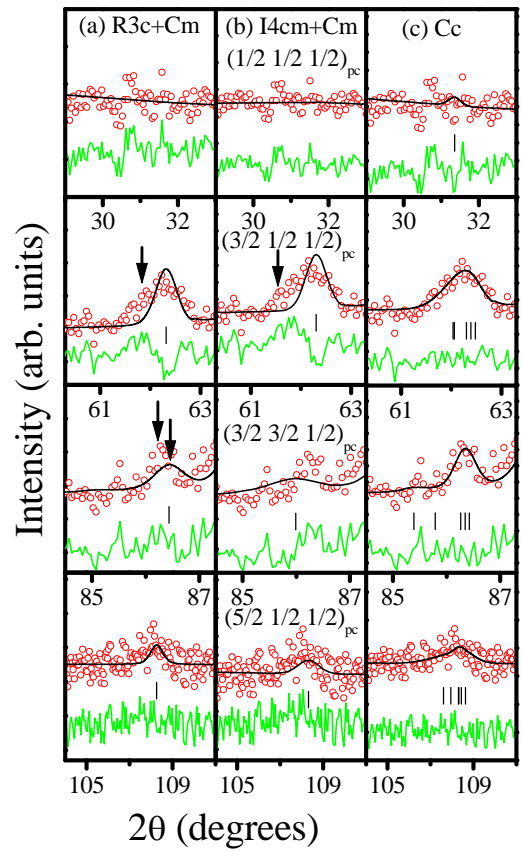

Fig. 5

# Supplementary File

**Table I.** Refined structural parameters and agreement factors for PZT525 at 4K for the R3c+Cm structural model

| Space group: R3c | | | | Space group: Cm | | | |
|---|---|---|---|---|---|---|---|
| Atoms  x | y | z | B (Å²) | x | y | z | B (Å²) |
| Pb  0.00 | 0.00 | 0.277(2) | β=0.228(7) | 0.00 | 0.00 | 0.00 | β=1.55(3) |
| Ti/Zr  0.00 | 0.00 | 0.018(4) | 0.001(1) | 0.467(5) | 0.00 | 0.567(6) | 0.009(4) |
| O1  0.130(2) | 0.349(3) | 0.0833 | 0.125(4) | 0.450(2) | 0.00 | 0.072(2) | 0.549(1) |
| O2 | | | | 0.200(1) | 0.2459(8) | 0.604(1) | 0.725(2) |

a= b= 5.6517(6)Å, c= 13.973(2)Å      a= 5.6749(2)Å, b= 5.6497(2)Å, c= 4.0302(2)Å

γ=120.00°                                                           β= 90.587°(2)

Weight fraction (%)               10.14                                              89.86

$\chi^2$=1.94, $R_P$=4.16, $R_{wp}$= 5.29, $R_{exp}$= 3.79